\documentclass{webofc}

\usepackage[varg]{txfonts}   
\usepackage{hyperref}
\usepackage{url}
\hypersetup{colorlinks=true,citecolor=blue,urlcolor=blue,linkcolor=blue}
\newcommand{\pfrac}[2]{\frac{\partial #1}{\partial #2}}
\newcommand{\lbk}{\left(}
\newcommand{\rbk}{\right)}

\begin{document}
\title{Effects of a background magnetic field on dilepton radiation}
%
%

\author{\firstname{Han} \lastname{Gao}\inst{1}\fnsep\thanks{\email{han.gao3@mail.mcgill.ca}} \and
        \firstname{Xiang-Yu} \lastname{Wu}\inst{1}\fnsep\thanks{\email{xiangyu.wu2@mail.mcgill.ca}} \and
        \firstname{Sangyong} \lastname{Jeon}\inst{1}\fnsep\thanks{\email{sangyong.jeon@mcgill.ca}} \and
        \firstname{Charles} \lastname{Gale}\inst{1}\fnsep\thanks{\email{charles.gale@mcgill.ca}}
        }

\institute{Department of Physics, McGill University, 3600 rue University, Montréal, QC, H3A 2T8, Canada}

\abstract{We explore the feasibility of using dilepton radiations as a QGP magnetometer. We calculate the dilepton production rate in the presence of a time-dependent magnetic field typically found in heavy-ion collisions. We compute the thermal dilepton spectra from Au+Au collisions  at  $\sqrt s$ = 19.6~GeV  BES energy— using a realistic (3+1)-dimensional multistage hydrodynamic simulation. We find the thermal dilepton elliptic flow to be very sensitive to the strength of the magnetic field.
}
\maketitle
\section{Introduction}
\label{intro}
In an off-central heavy-ion collision (HIC) event, two positive-charged colliding nuclei can create a magnetic field in the collision region where the quark-gluon plasma (QGP) forms. Estimations with the Biot-Savart law and detailed computations give a magnetic field strength of $eB \sim (1-10)m_\pi^2 \sim 10^{19}{\rm \ Gauss}$ for off-central HICs, which is among the strongest ones known in current universe~\cite{Huang:2015oca,Deng:2012pc}. Nevertheless, the field can experience a drastic decay during the pre-equlibrium stage of the QGP~\cite{McLerran:2013hla,Siddique:2025tzd}, resulting in a smaller QGP "effective" magnetic field. Experimentally, a measurement of $\Lambda$ hyperon polarization found the late-stage magnetic field to be \emph{at least} one order-of-magnitude lower than the Biot-Savart law estimation~\cite{STAR:2023nvo}.  
\par
If the effective magnetic field is weak in the QGP phase of the fireball evolution, treating it as a perturbation will be a proper approach. With this, we utilize the well-established and successful multi-stage hydrodynamic model of the QGP evolution as the background.
\par 
Electromagnetic (EM) probes including photons and dileptons, are powerful tool for direct extraction of many QGP properties such as temperature~\cite{Churchill:2023zkk} and electric conductivity~\cite{Rapp:2024grb}. Experimentally, EM probes have been recently measured at the Large Hadron Collider (LHC)~\cite{ALICE:2023jef} and at the Relativistic Heavy-Ion Collider (RHIC)~\cite{STAR:2023wta}. EM probes also couple to the external electromagnetic field, making them a potential candidate for quantifying the effective magnetic field of the QGP. In this study, we first calculate the leading-order EM-field-induced correction to the dilepton production. Combining with a realistic multi-stage hydrodynamic simulation, we provide predictions of the dilepton invariant mass spectrum and elliptic flow when a decaying background magnetic field $eB \sim (0.1-1)m_\pi^2$ is considered.

\section{Dilepton production rate}
\label{rate}
For simplicity, we consider only the Born rate of $q\bar q\to \ell^+\ell^-$, even though a more comprehensive dilepton production rate (DPR, dilepton differential yields per unit 4-volume) including the next-to-leading order QCD processesare available in ref.~\cite{Churchill:2023zkk}. The DPR for this annihilation process (in the limit of zero quark mass) is given by an integral over the quark phase space
\begin{equation}\label{eq:drd4k}
    \frac{dR}{d^4 k} =\sum_{q=u,d,s} \frac{\sigma(q\bar q\to \ell^+\ell^-) k^2}{2\pi^2} \int \frac{d^4 p_1}{(2\pi)^4} f_q(p_1) f_{\bar q}(k - p_1) \delta(p_1^2)\delta((k-p_1)^2).
\end{equation}
Here, $p_1$ is the quark 4-momentum, $k = p_1 + p_2 = p_+ + p_-$ is the 4-momentum for the lepton pair $\ell^+ \ell^-$, $\sigma(q\bar q\to \ell^+\ell^-)$ is the cross section of the indicated process and $f_{q/\bar q}$ is the distribution function of the quark/antiquark, respectively. 

In thermal equilibrium with a fluid velocity $u^\mu$, the quark (antiquark) distribution function $f_{q(\bar q)}$ takes the form of Fermi-Dirac function $n_F(u\cdot p) = \frac{1}{1+e^{(u\cdot p\mp \mu_q)/T}}$
where $T$ is the medium temperature and $\mu_q$ is the quark chemical potential\footnote{For a 3-flavour QGP, $\mu_q$ is linked with the baryon chemical potential $\mu_B$ by $\mu_q = \mu_B/3$}.
\par
In presence of an external EM field $F^{\mu\nu}$, the plasma is driven away from equilibrium. The non-equilibrium correction of the distribution function can be obtained by solving Boltzmann equation with the relaxation-time approximation~\cite{Sun:2023pil,Sun:2023rhh}
\begin{equation}
    p^\mu \partial_\mu f_q + e_q F^{\mu\nu}p_\mu \pfrac{f_q}{p^\nu} = - \frac{p \cdot u}{\tau_R} \delta f_q,
\end{equation}
where $\delta f(x,p) = f(x,p) - n_F(p)$ and $\tau_R$ is the relaxation time and $e_q$ is the electric charge of quark $q$. In the linear order of $F^{\mu\nu}$, one can obtain the correction to the quark distribution function as
\begin{equation}\label{eq:deltaf1}
    \delta f_q = \frac{e_q \tau_R}{T p \cdot u}n_F(p\cdot u - \mu_q)[1-n_F(p\cdot u - \mu_q)]p^\mu u^\nu F_{\mu\nu}.
\end{equation}
One can decompose the external $F^{\mu\nu}$ into in-fluid electric and magnetic field by fluid velocity $u^\mu$ as
\begin{equation}\label{eq:fmn}
    F^{\mu\nu} = E^\mu u^\nu - E^\nu u^\mu + \epsilon^{\mu\nu\rho\sigma}u_\rho B_\sigma,
\end{equation}
so eq.~(\ref{eq:deltaf1}) can be also written as
\begin{equation}\label{eq:deltaf2}
    \delta f_q = \frac{e_q \tau_R}{T p \cdot u}n_F(p\cdot u - \mu_q)[1-n_F(p\cdot u - \mu_q)]p^\mu E_\mu.
\end{equation}
\par 
The physics of eq.~(\ref{eq:deltaf2}) is understood as follows: in the rest frame of the fluid, even when the electric field and the magnetic field are both present, only the former is able to drive the plasma out-of-equilibrium. In the linear order, this is just the ordinary electric charge transport by Ohm's law $\vec j = \sigma_{\rm el} \vec E$. This observation allows us to match the relaxation time $\tau_R$ with the plasma conductivity $\sigma_{\rm el}$ by writing down the electric current given by eq.~(\ref{eq:deltaf2}), which gives
\begin{equation}\label{eq:sigma}
    \sigma_{\rm el} = \frac{4\pi \alpha_{\rm em} T^2}{9}\tau_R.
\end{equation}
\par
In the first order of $E^\mu$, eq.~(\ref{eq:drd4k}) reads
\begin{eqnarray}\label{eq:deltaR1}
\lbk \frac{d\delta R}{d^4 k}\rbk_{\rm EM} &=&\sum_{q = u,d,s}\frac{\sigma(q\bar q\to \ell^+\ell^-) k^2}{2\pi^2 T} \nonumber \\&\times &\int \frac{d^4 p_1 }{(2\pi)^4} \frac{e_q \tau_R p_1^\mu E_\mu}{u\cdot p_1}   n_F(u\cdot p_1 -\mu_q) \bar n_F(u\cdot p_1 -\mu_q) n_F(u\cdot(q-p_1) + \mu_q)  \nonumber\\ &+& (C\text{-conj.}).
\end{eqnarray}
Here, $C\text{-conj.}$ refers to the charge conjugation of the first part
, and we have also introduced a short-hand notation $\bar n_F(E) = 1-n_F(E)$.
As a scalar that depends on $u^\mu,k^\mu$ and $E^\mu$, the EM-field DPR correction must adapt the following form
\begin{equation} \label{eq:kde}
   \lbk \frac{d\delta R}{d^4 k}\rbk_{\rm EM} = \lbk b_0 u^\mu + b_1 \frac{k^\mu}{T}\rbk E_\mu = \frac{b_1}{T}k^\mu E_\mu.
\end{equation}
The factor $b_1$ can be worked out from eq.~(\ref{eq:deltaR1}) as
\begin{equation} \label{eq:b1}
    b_1 =\frac{3e\sigma_{\rm el} \sigma_{\rm tot} k^2 }{4(2\pi)^6 C_{\rm em}\alpha_{\rm em}T^2 |\vec k|^3}[2k_0T J_0(k_0/T,|\vec k|/T,\mu_q/T)- k^2 J_{-1}(k_0/T,|\vec k|/T,\mu_q/T) - (\mu_q \leftrightarrow -\mu_q)],
\end{equation}
where we introduced $\sigma_{\rm tot} = N_c C_{\rm em} \frac{16\pi \alpha_{\rm em}^2}{3q^2} B(m_\ell^2/q^2)$ as the total cross section of all $q\bar q\to \ell^+ \ell^-$ processes in the plasma with $N_c = 3$, $C_{\rm em} = \sum_{q=u,d,s} Q_q^2 = 2/3$ and the $J$ functions defined by
\begin{equation}
    J_n(a,b,z) = \int_{x_-}^{x_+} dx \frac{x^n}{(e^{x-z}+1)(e^{a-x+z}+1)(e^{z-x}+1)},\quad x_\pm = \frac{a\pm b}{2}.
\end{equation}

\section{Multi-stage hydrodynamic simulations}\label{hydro}
In this work, the dynamical evolution of the QGP medium is simulated using a (3+1)-dimensional multi-stage relativistic hydrodynamics framework~\cite{Schenke:2010nt,Schenke:2010rr} calibrated by the hadronic observables. 
Detailed descriptions of the model configuration and parameter calibration can be found Ref.~\cite{Du:2023gnv}.
\section{EM field profile}
In this work, we consider a decaying magnetic field in $-y$ direction in the lab frame as
\begin{equation}
    B_y(x,y,\tau) = -\rho(x,y)\frac{B_0}{1 + \tau/\tau_B},
\end{equation}
where $\rho(x,y)$ is a Gaussian smearing function and $\tau_B$ is the characteristic life time of the magnetic field. By Faraday's law, an electric field will be induced by the decay of the magnetic field, which is obtained from solving the Maxwell equation~\cite{Sun:2021joa}
\begin{equation}
    E_x(t,x,y,\eta_s) = \int_0^{\eta_s}d\chi\, \frac{\partial B(x,y,\tau)}{\partial \tau}\Bigg|_{\tau=\frac{t}{\cosh \chi}}\frac{t}{\cosh \chi}.
\end{equation}
Both the magnetic field $B_y$ and the induced electric field $E_x$ are collectively encoded into the lab-frame field tensor $F_{\mu\nu}$. Both fields contribute to the in-fluid electric field $E^\mu = F_{\mu\nu}u^\nu$ in the DPR correction eq.~(\ref{eq:kde}).
\section{Results}\label{results}
\begin{figure}[h]
\centering
\includegraphics[width=9cm,clip]{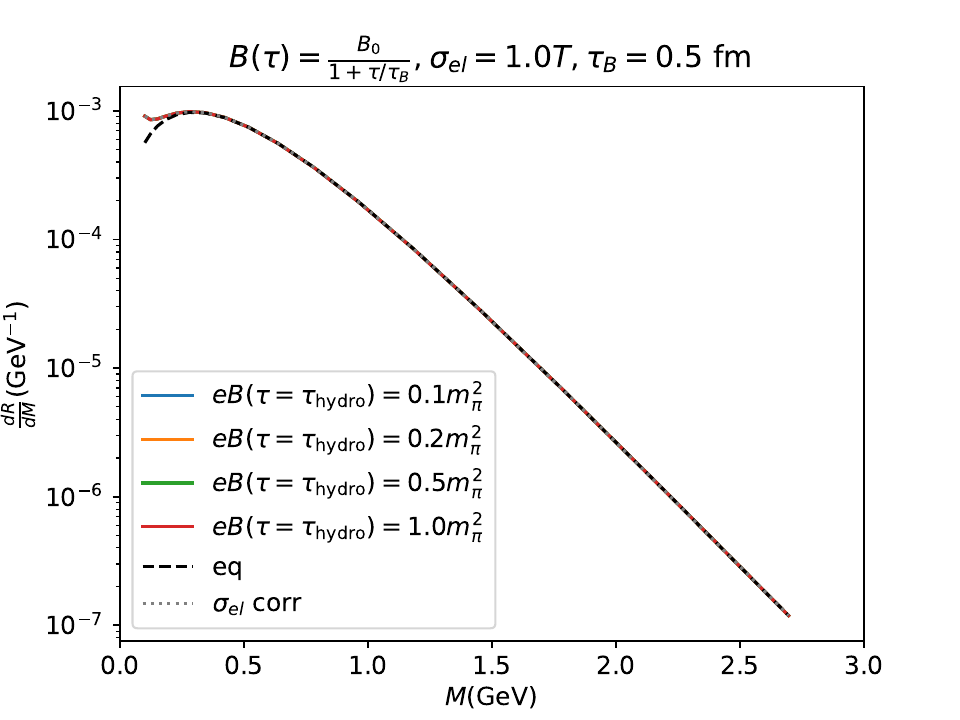}
\caption{Dilepton invariant mass spectrum $dN/dM$ for different values of $eB(\tau = \tau_{\rm hydro})$ and a fixed magnetic field lifetime $\tau_B = 0.5~{\rm fm}$. The black dash line represents the equilibrium result. The gray dotted lin includes the contribution from the relaxation channels responsible for a finite electric conductivity $\sigma_{\rm el}$, and the coloured solid lines further include the off-equilibrium effect caused by the external EM field, for different values of magnetic field at the starting time of the hydrodynamic phase. As these lines all overlap with the $\sigma_{\rm el}$ correction line with no magnetic field, we observe no correction from the external EM field.}
\label{fig:dndm}       
\end{figure}
Dilepton invariant mass spectra $\frac{dN}{dM}$ and elliptic flows $v_2(M)$ are calculated based on the multi-stage hydrodynamic simulation and the EM field profile, for the QGP phase of the fireball evolution, and with complimentary dilepton kinematics including transverse momentum integrated out. For simplicity, we keep a fixed electric conductivity parametrization $\sigma_{\rm el} = T$. The strength of the magnetic field at the beginning of the hydrodynamic evolution $B_h \equiv B(x=0,y=0,\tau = \tau_{\rm hydro})$ and the lifetime $\tau_B$ are chosen to be the free parameters. We also include the electric conductivity and the viscosity correction to the DPR~\cite{Vujanovic:2016anq}. Here, we present the results for AuAu collision at $\sqrt s = 19.6~{\rm GeV}$. 
\subsection{Dilepton invariant mass spectrum}
In fig.~\ref{fig:dndm}, we show the dilepton invariant mass spectrum for different values of $B_h$. No distinguishable difference is observed for $0.1m_\pi^2 < eB_h <1 m_\pi^2$. As a scalar, $dN/dM$ depends on the background magnetic field $\vec B$ as least quadratically, i.e., the lowest order of correction to $dN/dM$ is expected to come as $O(\vec B^2)$. Considering $\sigma_{\rm el}$ slightly increases dilepton yields in low-invariant-mass region. This observation is consistent with the NLO calculation~\cite{Churchill:2023zkk} , as a finite electric conductivity effectively includes relaxation channels beyond LO.
\begin{figure}[h]
\centering
\includegraphics[width=9cm,clip]{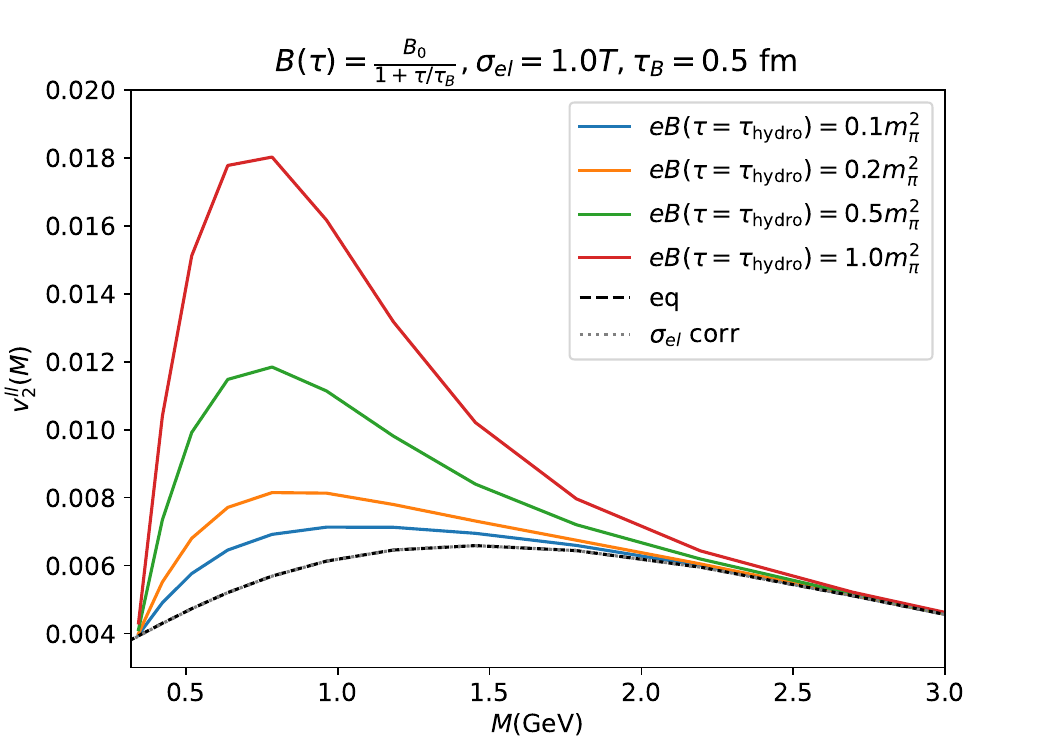}
\caption{Dilepton elliptic flow $v_2(M)$ for different values of $eB(\tau = \tau_{\rm hydro})$ and a fixed magnetic field lifetime $\tau_B = 0.5~{\rm fm}$. The colour code of the curves are the same as fig.~\ref{fig:dndm}.}
\label{fig:v2B}       
\end{figure}
\subsection{Dilepton elliptic flows}
In contrast to the case of invariant mass spectrum, in fig.~\ref{fig:v2B}, we observe a significant enhancement of the dilepton elliptic flow $v_2(M)$ when the strength of the background magnetic field increases, for $0.5~{\rm GeV}\lesssim M \lesssim 1.5~{\rm GeV}$.
\begin{figure}[h]
\centering
\includegraphics[width=9cm,clip]{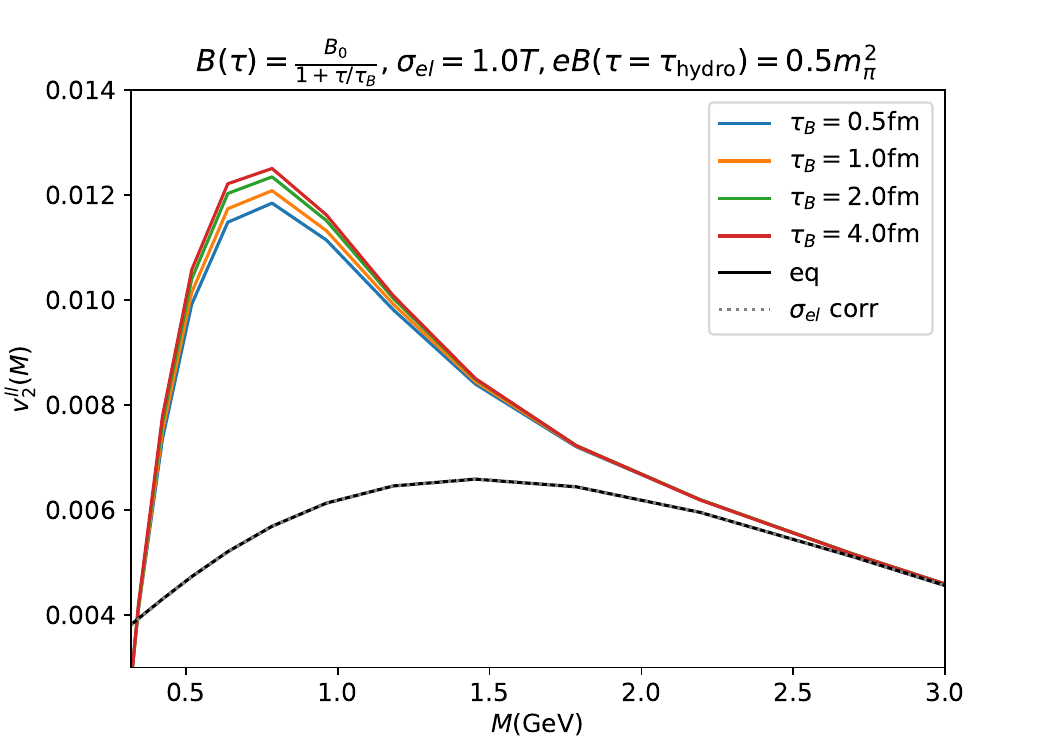}
\caption{Dilepton elliptic flow $v_2(M)$ for different values of magnetic field lifetime $\tau_B$ and a fixed $eB(\tau = \tau_{\rm hydro}) = 0.5m_\pi^2$. The curves are coulored differently for different value of $\tau_B$.}
\label{fig:v2T}       
\end{figure}
\par
In fig.~\ref{fig:v2T}, we keep a fixed value of $B_h$ and check the effect of different magnetic field lifetime $\tau_B$. One would first anticipate that a smaller $\tau_B$ means a faster decay of background EM field, so the average ``effective" EM field in the hydrodynamic phase is weaker, which finally results in a weaker $v_2$ signal. Instead, minuscular differences are observed between different $\tau_B$. This can be qualitatively understood by Faraday's law: a fast decaying magnetic field gives rise to a stronger electric field, which in return compensates the quick decay of the magnetic field. This makes dilepton $v_2$ a good probe for $B_h$: the strength of the background magnetic field at the beginning of the hydrodynamic evolution of the QGP.

\section{Summary}
In this work, we present a calculation of the non-equilibrium effect on dilepton production caused by an external EM field. For AuAu collision at 19.6~GeV, we found that a magnetic field $eB_h \sim (0.1-1)m_\pi^2$ at the starting time of the hydrodynamic evolution can give a significant enhancement of the dilepton elliptic flow $v_2$, which is also insensitive to the lifetime of the magnetic field. Experimentally measuring dilepton $v_2$ can therefore provide valuable insights of the background EM field of the hydrodynamic phase of the fireball evolution.
\par
\vspace{0.5cm}
\noindent\textbf{Acknowledgements:} The authors thank Jing-An Sun for many helpful discussions. This work was supported in part by the Natural Sciences and Engineering Research Council of Canada. Computations were made on the B\'eluga and Narval computers managed by Calcul Qu\'ebec and by the Digital Research Alliance of Canada.
%
%

\end{document}